\documentclass[12pt,preprint]{aastex}

\shorttitle{High Redshift X-ray selected QSOs}
\shortauthors{Treister et al.}

\begin{document}

\title{An X-ray Selected AGN at z=4.6 Discovered by the CYDER Survey\footnote{Partly based on observations collected at the
European Southern Observatory, Chile, under program 72.A-0509}}

\author{Ezequiel Treister\altaffilmark{2,3},Francisco J. Castander\altaffilmark{4},Thomas J. Maccarone\altaffilmark{5},
 David Herrera\altaffilmark{2},Eric Gawiser\altaffilmark{3,2,6}, Jos\'e Maza\altaffilmark{2} and Paolo S. Coppi\altaffilmark{1} }
\email{treister@astro.yale.edu}

\altaffiltext{2}{Department of Astronomy, Yale University, P.O. Box 208101,New Haven,CT 06520.}
\altaffiltext{3}{Departamento de Astronom\'{\i}a, Universidad de Chile, Casilla 36-D, Santiago, Chile.}
\altaffiltext{4}{Institut d'Estudis Espacials de Catalunya/CSIC,Gran Capit\`a 2-4, E-08034
Barcelona, Spain.}
\altaffiltext{5}{Astronomical Institute ``Anton Pannekoek'',University of Amsterdam,1098 SJ Amsterdam, The Netherlands}
\altaffiltext{6}{NSF Astronomy and Astrophysics Postdoctoral Fellow}

\begin{abstract}
We present the discovery of a high redshift, X-ray selected AGN by the
Calan-Yale Deep Extragalactic Research (CYDER) survey: CXOCY
J033716.7-050153, located at $z=4.61$, the second high redshift AGN
discovered by this survey. Here, we present its optical, near-IR and
X-ray properties and compare it with other optical and X-ray selected
high redshift AGN. The optical luminosity of this object is
significantly lower than most optically selected high redshift
quasars. It also has a lower rest frame UV to X-ray emission ratio
than most known quasars at this redshift. This mild deviation can be
explained either by dust obscuring the UV radiation of a normal radio
quiet AGN emitting at $10\%$ of its Eddington luminosity or because
this is intrinsically a low luminosity radio loud AGN, with a
super-massive Black Hole of $\sim 10^8M_\sun$ emitting at 1\% of its
Eddington luminosity. Deep radio observations can discriminate between
these two hypotheses.
\end{abstract}
\keywords{galaxies: active --- quasars: individual (CXOCY J033716.7-050153) --- X-rays: galaxies}


\section{Introduction}

Due to its large intrinsic luminosity, an AGN (Active Galactic
Nucleus) can be observed up to very high redshifts, and AGNs can
therefore be used to probe the early epochs of the universe when the
formation of large structures began. Recent observations have found a
considerable number of quasars at high redshift
\citep{anderson01,schneider02}, with some of them at redshift $>6$
\citep{fan03}, when the universe was about 1 Gyr old.

While AGNs were first noticed to be a class of objects worthy of
detailed follow-up in other bands on the basis of their radio
emission, e.g. \citep{schmidt63}, more recent work has focused on
surveys at shorter wavelengths.  Due to their low spatial density,
most AGNs at high redshift (defined here to be $z>4$) were discovered
by large area, shallow optical surveys. An example of this is the
Sloan Digital Sky Survey (SDSS; \citealt{york00}), which in its first
data release \citep{schneider03} presented 238 high redshift
quasars. However, these quasars were selected based on their optical
properties, and optical surveys are subject to bias against the
discovery of quasars which are either intrinsically dim at these
wavelengths or are obscured by dust like the type 2 AGN that are
believed to comprise most of the AGN background \citep{hasinger02}.

The other main AGN selection technique is X-ray detection, as X-ray
emission appears to be a universal characteristic of AGNs at all
observed redshifts \citep{kaspi00}. However, given the relative
faintness of the majority of the sources, X-ray surveys need to rely
on optical spectroscopy for most of the identifications. With new
X-ray missions like the Chandra X-ray Observatory and its superb
angular resolution, the identification of X-ray sources at other
wavelengths has become much easier, allowing for successful
multiwavelength follow-up of X-ray sources. The nature of the X-ray
emission also ensures that the optical selection bias due to
obscuration is strongly reduced (although the X-ray selection may
introduce other biases). This is especially true at high redshift,
where the observed X-ray photons were emitted at higher energies and
can penetrate even considerable amounts of obscuring material.

According to a recent review by \citet{brandt03} only seven of the high redshift AGNs known were discovered by their
X-ray emission. Three of them were found with ROSAT  \citep{henry94,zickgraf97,schneider98}, while the other 
four were detected by Chandra \citep{silverman02,barger02,castander03a}. One of them was detected in the radio and
classified as radio loud \citep{zickgraf97}.

In this paper, we present the discovery of the eighth high redshift
X-ray selected AGN, and the second discovered by the Calan Yale Deep
Extragalactic Research (CYDER) survey, at a redshift $z=4.61$. The
CYDER survey is a collaborative effort between Universidad de Chile
and Yale University to study faint stellar and extragalactic
populations in detail. One of the key aims is the characterization of
the population of faint X-ray sources. In order to do this, some of
the fields of the survey were selected to overlap with moderately deep
($\sim$ 50 ks) Chandra pointings. Multiwavelength follow-up in these
fields includes optical imaging using the CTIO 4m telescope, near IR
images obtained using the DuPont telescope at Las Campanas Observatory
(LCO) and optical spectroscopy from VLT at Cerro Paranal and Magellan
at LCO. A more complete review of the optical and spectroscopic
properties of the X-ray sources in the first two fields can be found
in \citet{castander03b}.


In \S 2 we describe the X-ray, optical and near-IR observations that
led to the discovery of the AGN presented here.  In \S 3 we discuss
its observed properties and compare it with other high redshift
quasars, both optical and X-ray selected.  We present our conclusions
in \S 4. Throughout this paper we assume $H_0=70h_{70}$ km s $^{-1}$
Mpc$^{-1}$, $\Omega_m=0.3$ and $\Omega_\Lambda =0.7$. These
cosmological parameters are consistent with the recent findings by
\citet{spergel03}.  We define the photon index $\Gamma$ as the
exponent giving a photon flux density X-ray spectrum $dN/dE\propto
E^{-\Gamma}$ in photons cm$^{-2}$ s$^{-1}$ keV$^{-1}$.

\section{Observations and Data Analysis}

\subsection{X-Ray Data}

As part of the CYDER Survey, five fields observed by Chandra and
currently available in the archive were downloaded and analyzed using
standard techniques with the CIAO package. In this paper we discuss
the results for the source CXOCY J033716.7-050153 found in the SBS
0335-05 field. This field was observed by Chandra on September 7th,
2000 for 60.51 ks (PI: Thuan) and retrieved from the archive by our
group. A detailed analysis of the whole sample of X-ray sources
detected in these fields will be presented in \citet{treister03}.

Reduction of the data included the removal of bad columns and pixels
using the guidelines specified on the ``ACIS Recipes: Clean the Data''
web page and the removal of flaring pixels using the FLAGFLARE
routine. We used the full set of standard event grades (0,2,3,4,6) and
created two images, one from 0.5 to 2.0 keV and one from 2.0 to 8.0
keV. Then, we used the WAVDETECT routine from the CIAO package to
identify the point sources within these images, checking wavelet
scales 1,2,4,8 and 16.

CXOCY J033716.7-050153 was detected in the I2 ACIS CCD. We extract an
X-ray spectrum using a minimum of 5 counts per channel from 0.5 to 3.5
keV (i.e. the 20 lowest energy photons), using the PSEXTRACT script.
We then fitted the spectrum within XSPEC 11.0, with a power law model
and absorption constrained to the Galactic value of $4.8\times10^{20}$
from the $N_H$ tool.  This is relatively unconstraining, yielding a
spectral index of $\Gamma=2.0\pm0.6$, and observed fluxes of
$1.5\times10^{-15}$ ergs/sec/cm$^2$ and $3.2\times10^{-15}$
ergs/sec/cm$^2$, from 0.5-2 and 0.5-8 keV, respectively. The fit to
the X-ray spectrum did not depend on whether the $\chi^2$ or the
C-statistic was used, so it is unlikely that large errors were induced
by fitting with an insufficient number of photons per bin. The
unabsorbed fluxes are approximately $1.8\times10^{-15}$ and
$3.5\times10^{-15}$ ergs/sec/cm$^2$, respectively.  The errors should
be about 30\% due to the statistical error in the count rate and the
uncertainties in the best-fitting spectral models. The cumulative
histogram of detected X-ray photons is presented on Figure 1.

\subsection{Optical Imaging Data}

The field of the AGN was observed with the CTIO 4 meter Blanco
telescope, using the MOSAIC-II camera, that provides a field of view
of 36$'$x36$'$. Images were taken using the $B$,$V$,$R$ and $I$
filters on the nights of October 8th and 10th, 2002, centered on the
Chandra archival image central coordinates. Exposure times were 2
hours under 1.3$''$ seeing conditions in the $B$ filter, 50 minutes
under 1.1$''$ seeing in $V$, 50 minutes under 1.1$''$ seeing in $R$
and 20 minutes under 1.0$''$ seeing in $I$.

A second image in the $I$ band was taken with the UT4 VLT telescope at
Cerro Paranal, Chile using the FORS2 instrument. This image was taken
on the night of September 19 2003 in service mode. Total exposure time
was 14 minutes under 0.7$''$ seeing conditions.

Images were reduced using standard techniques with the IRAF/MSCRED
package. Identification of the optical counterpart of the X-ray source
was straightforward given the superb spatial resolution of the Chandra
Observatory. Indeed, only one optical source can be found in a 3$''$
circle around the X-ray emission centroid. The offset between the
X-ray and optical positions was $<0.1''$ in RA and $0.5''$ in Dec.  In
order to calculate the received flux in each band, we performed
aperture photometry using an aperture of $1.4\times$FWHM of each
image, centered on the centroid of the $R$ band source, the band in
which the signal to noise is highest. The observed magnitudes in each
band can be found in Table 1. Figure 2 shows the optical finding
chart, based on the $R$ band image.

\subsection{Optical Spectroscopy}

Multislit spectroscopy of X-ray sources detected by Chandra in the
field of the AGN was obtained at Las Campanas Observatory with the
Magellan I (Baade) telescope, using the LDSS-2 instrument. The
observations were taken on the night of October 4th, 2002. The field
of the high redshift AGN CXOCY J033716.7-050153 was observed for 2
hours in 0.75$''$ seeing conditions. The Med/Blue grism was used,
giving a dispersion of 5.3\AA~ per pixel at a central wavelength of
5500\AA. The obtained spectrum was reduced using standard IRAF tasks
called by a customized version of the BOGUS code\footnote{Available at
http://zwolfkinder.jpl.nasa.gov/$\sim$stern/homepage/bogus.html} .We
calibrated the wavelength of the spectrum using the He-Ar comparison
lamp and the night sky lines. Rough flux calibration of the spectrum
was performed using the spectrum of the LTT9239 spectrophotometric
standard.

Figure 3 shows the final, reduced spectrum together with the mean
spectrum of quasars at $z>4$ observed by the SDSS. Ly$\alpha$ and NV
are clearly distinguishable in the spectrum. The measured redshift
using these emission lines is $z=4.61\pm 0.01$.

\section{Discussion}

With an absolute magnitude of $M_B=-21.14$ (Vega), calculated
extrapolating the $I$ band magnitude using the SDSS QSO composite
spectrum \citep{vandenberk01}, CXOCY J033716.7-050153 is a faint
AGN. Indeed, there are only two AGNs known at high redshift that are
fainter in the optical, VLA J1236+6213 at $z=4.42$ \citep{brandt01}
and CXOHDFN J123719.0+621025 at $z=4.13$ \citep{barger02}, both
located in the HDF-N/CDF-N.  In X-rays, this quasar shows a luminosity
in the [0.5-2.0] keV band of $3.77 \times 10^{44}$ ergs s$^{-1}$. This
is a large X-ray luminosity for an AGN that is faint in the optical
bands. In order to quantify this statement, we can calculate the
effective optical to X-ray power-law spectral slope which is given by
\begin{equation}
\alpha_{ox}=\frac{\log [f_\nu (2\textnormal{ keV})/f_\nu (2500\textnormal{ \AA}]]}{\log [\nu (2\textnormal{ keV})/\nu (2500\textnormal{ \AA})]}
\end{equation}
where $f_\nu$ is the flux density per unit of frequency and $\nu$ is
the frequency of the given wavelength or energy. For this AGN, we
measure a value of $\alpha_{ox}=-1.16\pm 0.16$. The average value for
the known sample of $z>4$ AGNs detected in X-rays\footnote{This sample
can be found in the Web site
http://www.astro.psu.edu/user/niel/papers/highz-xray-detected.dat
maintained by Niel Brandt and Christian Vignali} is $-1.61\pm 0.24$
from a total of 71 sources, so this new AGN is a $\sim 2\sigma$
deviation. With the data available now we cannot calculate the
contribution from the host galaxy to the optical luminosity, but it
can only be significant if its luminosity is $\ga 3L_*$. Furthermore,
a significant contribution from the host galaxy will only make the
optical luminosity of the AGN to be smaller, therefore increasing the
value of $\alpha_{ox}$ and making the deviation from a typical AGN at
this redshift to be even larger.

A plot of $\alpha_{ox}$ as a function of redshift for all
the high redshift AGNs detected in X-rays can be found in
Figure 4 and the $F_X/F_{UV}$ relation for these objects is
presented in Figure 5. In Figure 4 we can see that the X-ray
selected sources have systematically higher values of
$\alpha_{ox}$ than optically selected AGNs. This correlation
is expected from selection effects; however, it should be
noted that the optical luminosities of all the X-ray
selected sources are fainter than that of the least luminous
optically-selected quasar (with the exception of the one
X-ray selected blazar), see Figure 5.

Recents results presented by \citet{vignali03b} show that
there is a good correlation between the UV flux
characterized by the flux density at 2500\AA~ and the soft
X-ray flux measured at 2 keV. This correlation can be
expressed as $F_X\propto F_{UV}^{0.75}$. A similar result
was reported in \citep{castander03a}, in this case
$F_X\propto F_{UV}^{0.55}$. This relation implies that
$\alpha_{ox}$ should also be correlated with the UV flux (or
luminosity). Figure 5 shows that optically selected AGNs are
more luminous, both in the optical and the X-ray, than the
X-ray selected ones. The $\alpha_{ox}$ value of optically
selected AGN are therefore expected to be lower as can be
seen in Figure 4.


In Figure 5, we can also see that without correcting for the
possible optical obscuration CXOCY J033716.7-050153 is a
significant deviant from the $F_X-F_{UV}$ relation. There
are two possible explanations for this deviation: either
there is significant obscuration of the optical light by
dust or this a radio loud AGN with $\alpha_{ox}$ enhanced by
a second component, probably coronal, contributing to the
X-ray emission. Strong optical variability can be ruled out
using the two optical observations separated by $\sim 1$
year in the observed frame, in which the $I$ band magnitude
remained almost unchanged given the error bars.

If this AGN is relatively unobscured, then its anomalous value of
$\alpha_{ox}$, as well is its relative optical faintess compared with
other high redshift AGN may be explained if the source is at the
bright end of the low luminosity AGN (LLAGN) category.  Such sources
are typically seen to have X-ray luminosities below about 1\% of the
Eddington limit, e.g. \citet{ho02,maccarone03}.  The mass of the
central black hole would then have to be at least about $3.4\times10^8
M_\sun$.  Applying the black hole fundamental plane relation from
\citet{merloni03},
\begin{equation}
\log L_R =(0.60^{+0.11}_{-0.11})\log L_X+(0.78^{+0.11}_{-0.09})\log M+7.33^{+4.05}_{-4.07}
\end{equation}
where $L_R$ is the radio luminosity at 5 GHz, $L_X$ is the 2-10 keV
X-ray luminosity and M is the mass of the black hole in units of
$M_\sun$, we find that the radio power is expected to be at least
$f_{5 GHz}=31.5\mu$Jy. Now, if we assume instead that this is a
typical AGN radiating at about 10\% of its Eddington luminosity then
using the Merloni relation we get a radio flux of $f_{5
GHz}=5.2\mu$Jy. Moreover it seems that the radio emission is quenched
in AGN accreting at about 10\% of the Eddington rate
\citep{maccarone03}, so an even lower radio flux would be expected.

Following the standard definition of radio to optical flux
ratio $R_{ro}$ given by \citet{kellermann89}, where optical
flux is defined as the flux at 4400\AA~ in the rest frame
and radio flux is calculated at a rest frame frequency of
5 GHz (6 cm), radio loud sources have $R_{ro}$ values in the
range 10-1000, while for radio quiet AGNs
0.1$<R_{ro}<$1. With the assumption that the Eddington ratio
is 1\% for this AGN we get that $R_{ro}=15.06$, making it a
radio loud AGN. However, given the low optical luminosity,
we cannot neglect the contribution of star light from the
host galaxy, in which case the value of $R_{ro}$ computed is
simply the lower limit of the radio to optical flux ratio
for the AGN emission alone. Assuming an Eddington ratio of
10\% then $R_{ro}=2.49$, which makes it a radio quiet AGN.

According to an alternate definition of radio loudness often
used for high redshift quasars \citep{stern00}, to be
considered radio loud an AGN has to have a minimum 1.4 GHz
specific luminosity of $L_{\textnormal{1.4 GHz}}=1.61\times
10^{32}$ ergs s$^{-1}$ Hz$^{-1}$
\citep{schneider92}, which at $z=4.61$ corresponds to a specific flux
of $f_{\textnormal{1.4 GHz}}=0.9$ mJy, very close to the detection
limit of the VLA FIRST survey. Unfortunately, the field of CXOCY
J033716.7-050153 was not observed by the FIRST survey. Using the NVSS
survey \citep{condon98} catalog, there were no detectable sources
within a 1 arcminute radius. Considering that the completeness limit
is 2.5 mJy, we can use this as an upper limit to the 1.4 GHz continuum
emission. Then, this AGN could be at most marginally classified as
radio loud.

In order to determine the amount of obscuration affecting the optical
spectrum of this AGN we can compare its optical colors with a typical
AGN spectrum at this redshift. The optical colors of this AGN are:
$V-R=1.75\pm 0.40$ and $V-I=2.1\pm 0.39$. Using the composite SDSS QSO
spectrum \citep{vandenberk01} and redshifting it to $z=4.61$ we obtain
the following colors: $V-R=1.02$ and $V-I=1.71$ accounting for the IGM
absorption using the description given in \citet{madau95}. If instead
we use an average of all SDSS QSOs at $z>4$ the results are very
similar, finding differences $\la 10\%$. The fact that CXOCY
J033716.7-050153 is slightly redder than the average QSO at that
redshift suggests that the optical spectrum is subject to some
obscuration in the line of sight. In order to investigate this
hypothesis, we added dust obscuration to the average spectrum
following the prescription given in \citet{cardelli89}. Given that the
observed $R$ band flux is affected by the Ly$\alpha$ emission line,
whose strength is highly variable from AGN to AGN, we will base our
reddening estimate only on the $V-I$ color. If order to obtain a $V-I$
color of 2.1 for the redshifted composite SDSS quasar, similar to the
observed value we need to add an extinction contribution of
$A_V=0.41\pm 0.4$ in the rest frame. Therefore, we will adopt this
value to correct the observed optical fluxes for intrinsic dust
extinction. It is also worth noting that the accretion disk may become
cooler at lower luminosities, which would also yield redder colors.

Another typical way to calculate the amount of obscuration in the line
of sight is using the X-ray spectrum. In this case, however, we do not
have enough counts in the X-ray spectrum to calculate absorption
directly. Also, since the observed soft X-ray band [0.5-2.0] keV
translates into [2.8-11.22] keV in the rest frame, the observed X-ray
spectrum is insensitive to moderate amounts of neutral hydrogen
absorption ($N_H<10^{23}$ cm$^{-2}$). Using the standard dust to gas
ratio, we can convert the optical extinction to a neutral hydrogen
column density, given by $N_H=1.96\times 10^{21}A_V$ cm$^{-2}$
\citep{granato94}. Then, we obtain in this case that $N_H=8\times
10^{20}$ cm$^{-2}$, implying that the X-ray spectrum is not affected
by absorption in the observed energy range.

Now, correcting for the optical obscuration we compute an unobscured
$I$ magnitude of 22.52$\pm 1.1$. Using this value, we calculate the
unobscured optical to X-ray power law slope $\alpha_{ox}=-1.35$, which
is now similar to the average value found for high redshift AGN.

As we can see in Figure 5 and already presented in
\citet{castander03a}, $f_{X}\propto f_{UV}^{0.55}$ is a better fit to
the data if the low UV luminosity X-ray selected AGNs are
included. The values for CXOCY J033716.7-050153 are consistent with
this correlation if optical obscuration is taken into account.

\section{Conclusions}

We present the discovery of the CXOCY J033716.7-050153, the second
high redshift X-ray selected AGN discovered by the CYDER survey.  This
is a faint object in the optical. In fact there are only two AGNs fainter
at these high redshifts. The X-ray flux relative to its optical
emission is unusual, having a larger value of $\alpha_{ox}$ than the
typical AGN at high redshift. There are two effects that could bring 
this object into better accordance with the observed correlations
for the general high redshift population:

The first possibility is that this is a radio loud low luminosity AGN
with a Black Hole of $\sim 3\times 10^8M_\sun$ , emitting at $\sim
1\%$ of its Eddington Luminosity. In this case the accretion disk is
cooler than a typical AGN at high redshift, making the optical colors
look redder.  Also, the high value of $\alpha_{ox}$, even taking into
account the correlation with optical luminosity, can be explained by
the presence of second emission component, probably coronal, which
contributes to the X-ray flux measurement.

A second possibility is that this is a radio quiet AGN emitting at an
Eddington ratio of $\sim 10\%$, in which the optical emisssion is
obscured by a moderate amount of dust which would imply a neutral
hydrogen column density of $\sim 10^{21}$ cm$^{-2}$, causing the
observed optical colors to be redder than the normal AGN. If this is
the case, the observed value of $\alpha_{ox}$ is affected by the
obscuration of the optical light, and after correcting for it, the
corrected value agrees very well with the measurements for other high
redshift AGNs, including the observed correlation between $f_{X}$ and
$f_{UV}$.
 
Detection of a flat radio spectrum in this source above about 10
$\mu$Jy would confirm the Low Luminosity AGN hypothesis, while a
non-detection in the radio with a sensitivity limit of a few $\mu$Jy
would strongly support the radio quiet, mildly obscured AGN
hypothesis.

\acknowledgments

We like to thank Meg Urry for useful discussions and the anonymous
referee for his comments. The CYDER survey participants and especially
FJC, ET and EG acknowledge support from Fundacion Andes. JM gratefully
acknowledges support from the Chilean Centro de Astrof\'\i sica FONDAP
15010003. EG acknowledges the support of an NSF Astronomy \&
Astrophysics Postdoctoral Fellowship under award AST 02-01667.

\clearpage

\begin{figure}
\figurenum{1}
\label{fig1}
\plotone{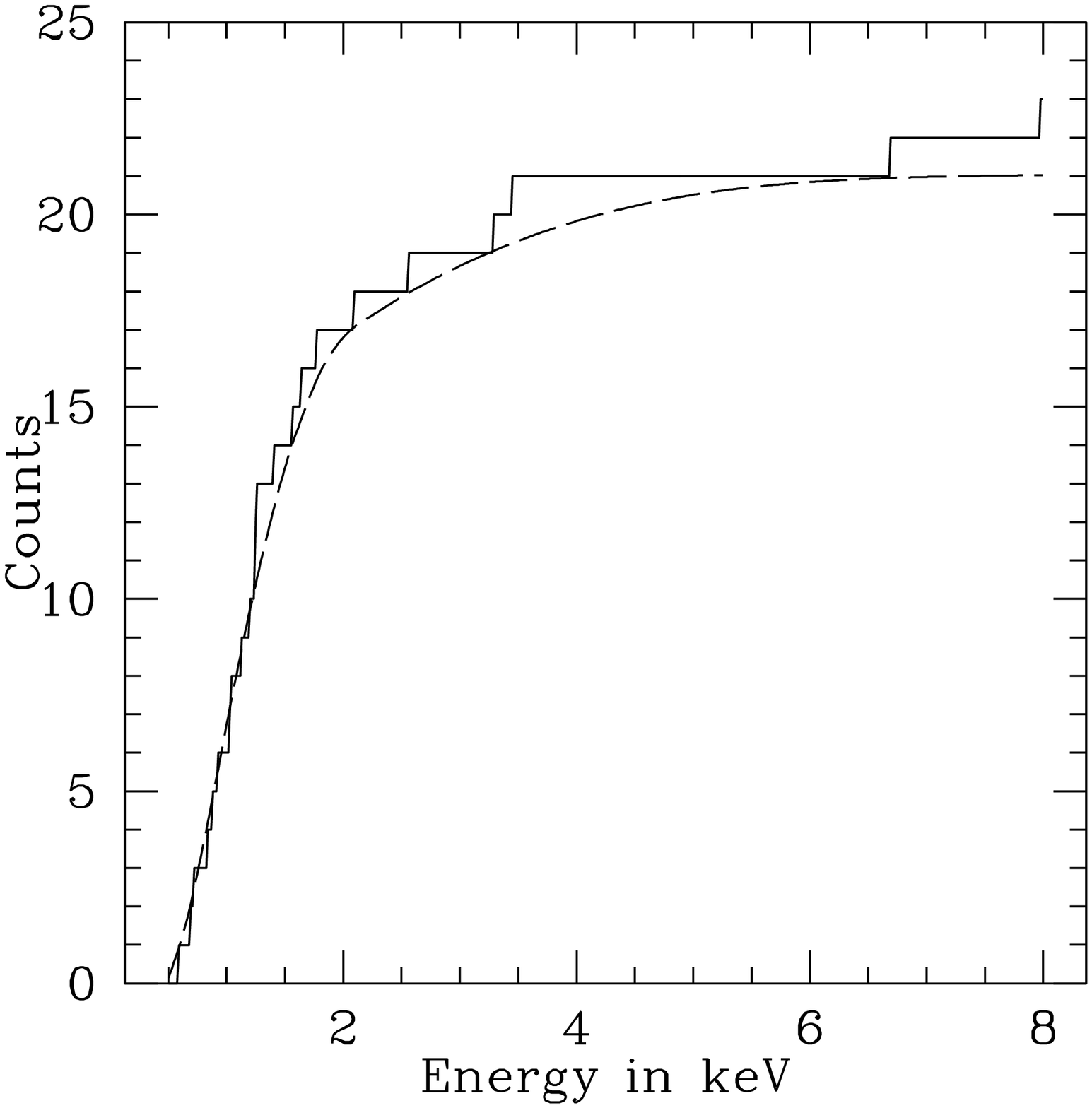}
\caption{Cumulative histogram of detected X-ray photons. The solid line shows the distribution of the data, while the dashed line
shows the fitted model with $N_H=4.8\times 10^{20}$ cm$^{-2}$ and $\Gamma=2.0$. Errors in the measurement are Poissonian.}
\end{figure}

\begin{figure}
\figurenum{2}
\label{fig2}
\plotone{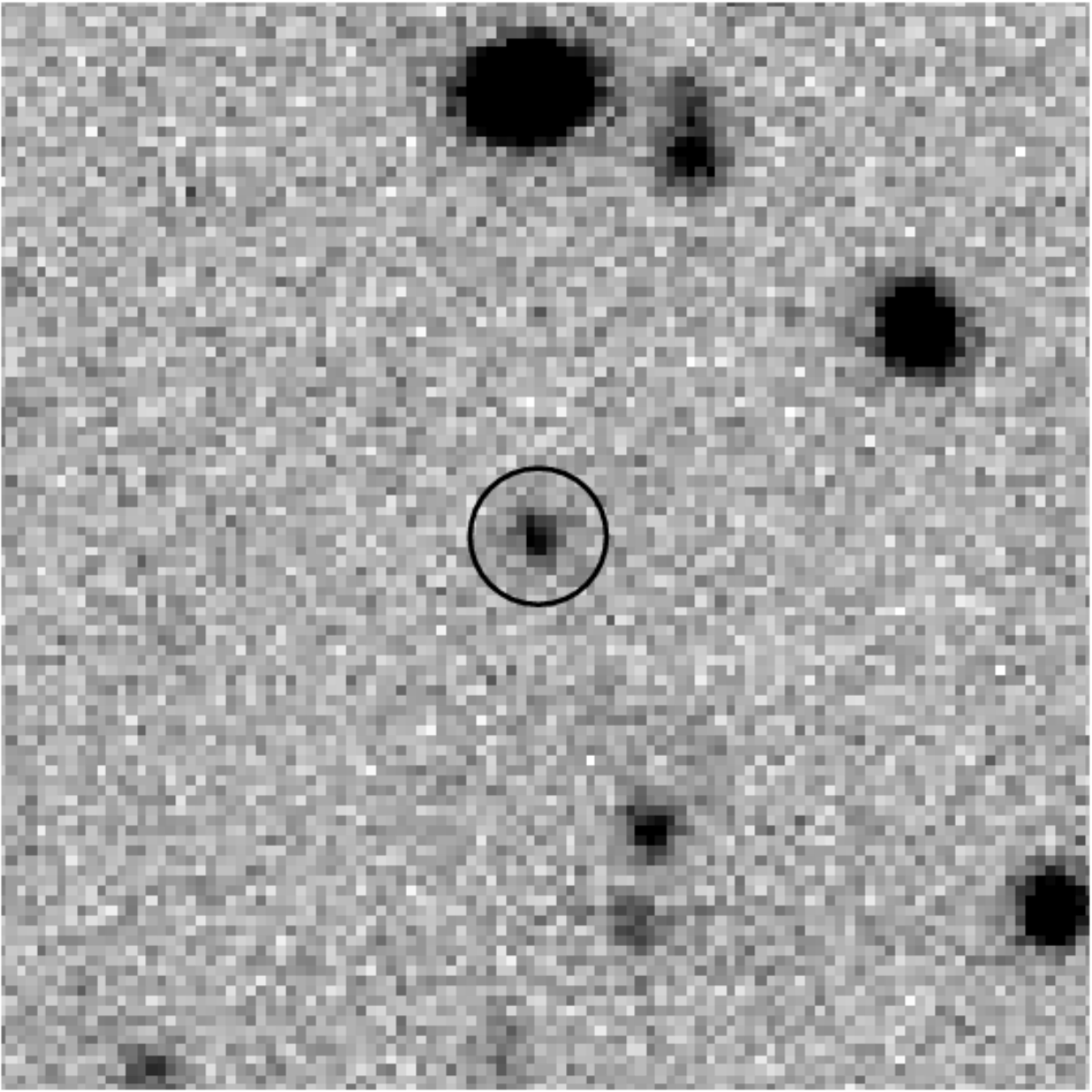}
\caption{R band image of CXOCY J033716.7-050153. This image is a cutout of 30$''$x30$''$ of the original image, centered on the position
of the optical counterpart of the X-ray source. North is up and east is to the left. This image was taken at the 
CTIO 4m telescope using the MOSAIC-II camera. A 2$''$ circle centered on the optical counterpart position is also plotted.}
\end{figure}

\begin{figure}
\figurenum{3}
\label{fig3}
\plotone{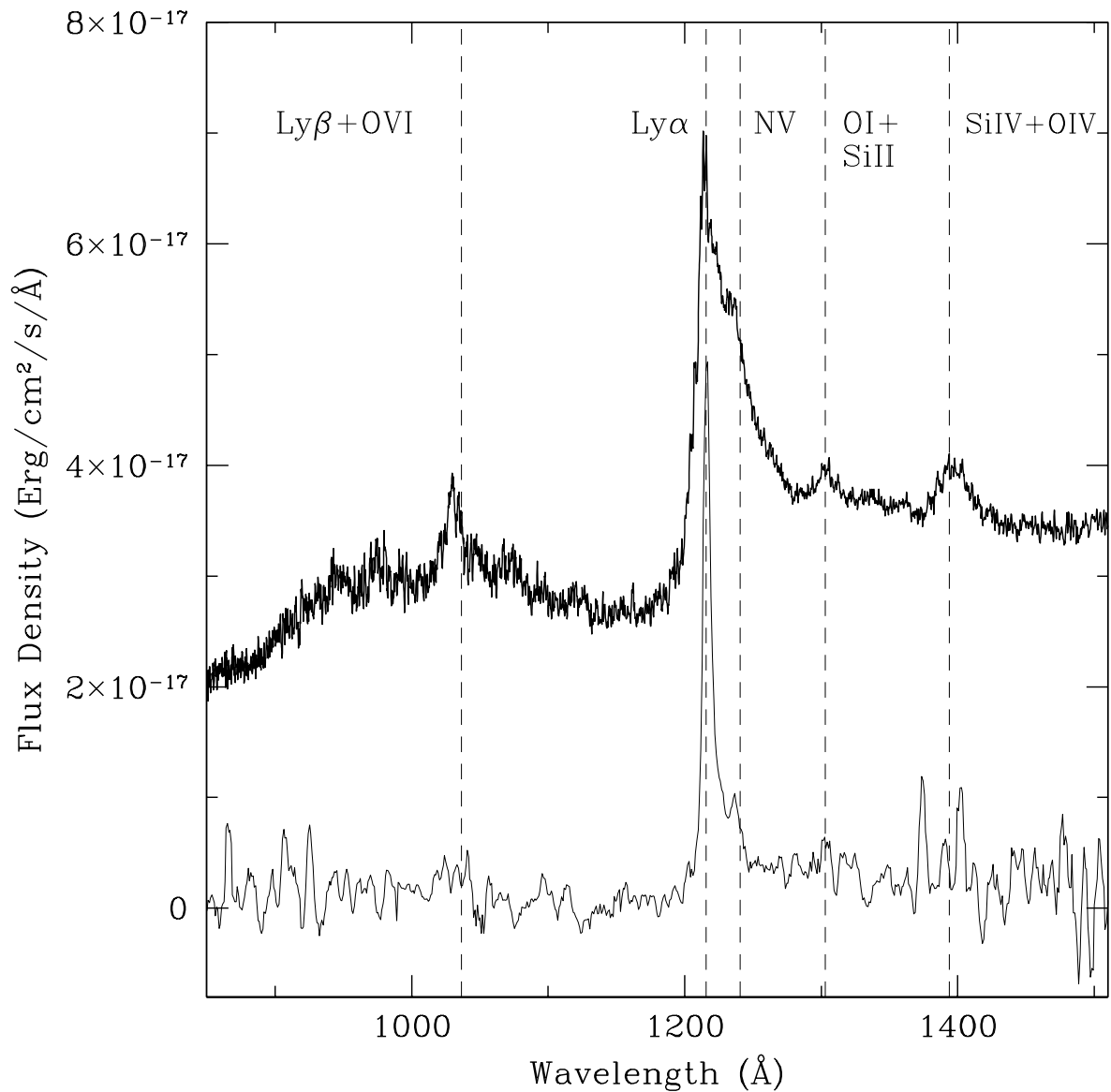}
\caption{Observed optical spectrum of CXOCY J033716.7-050153 shifted into the rest frame using the best fit redshift
of 4.61.
Original dispersion is 5.3\AA~ per pixel. The spectrum shown was boxcar smoothed using a 5 pixel box. Resolution of the
rest frame smoothed spectrum is 5.6\AA~.
For comparison we also show the error-weighted average of the SDSS Early Data Release QSO spectra at $z>4$. See 
\citep{castander03a} for details.}
\end{figure}

\begin{figure}
\figurenum{4}
\label{fig4}
\plotone{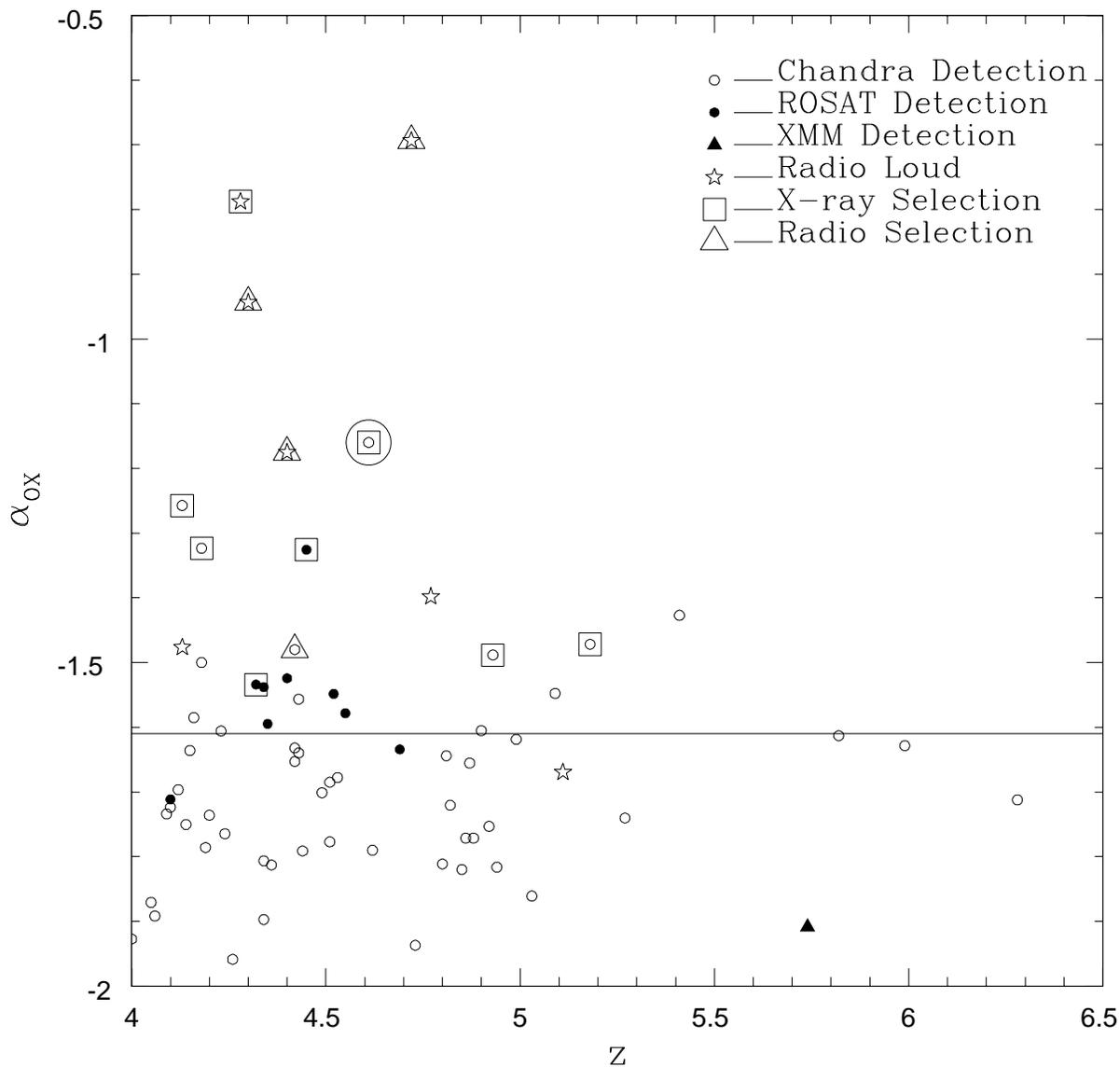}
\caption{Plot of $\alpha_{ox}$ vs. redshift for the known sample of high redshift AGN detected in X-rays. Filled circles are
ROSAT detections, empty circles are Chandra detections and the filled triangle is the only XMM-Newton detection. Stars are
radio loud AGNs. Symbols enclosed by squares denote X-ray selection, those enclosed by triangles denote radio selection, while 
no enclosing sign denotes 
optical selection. The horizontal line marks the position of $\alpha_{ox}=-1.61$, the average of the $\alpha_{ox}$
distribution. The large circle encloses the position of CXOCY J033716.7-050153.}
\end{figure}

\begin{figure}
\figurenum{5}
\label{fig5}
\plotone{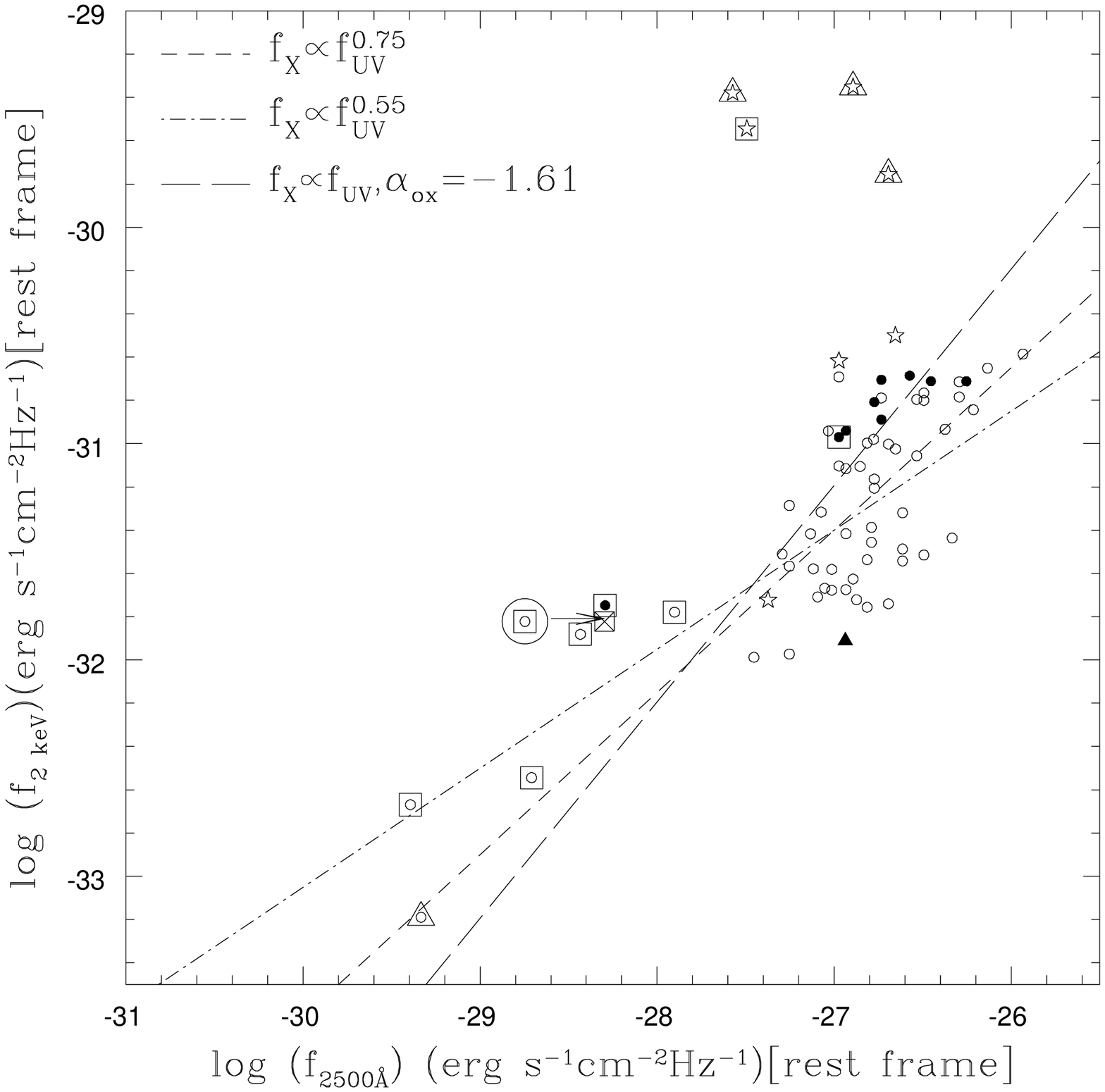}
\caption{$f_{\textnormal{2k}}$ vs $f_{\textnormal{2500 \AA}}$ (as defined in the text) for the known sample of high redshift AGN detected in X-rays.
Symbols are the same as in Fig 4. Segmented lines show the positions of the $f_X\propto f_{UV}^{0.75}$ and
$f_X\propto f_{UV}^{0.55}$ correlations reported by \citep{vignali03b} and \citep{castander03a} respectively. 
The long-dashed line shows the position of sources with $f_X\propto f_{UV}$ and $\alpha_{ox}=-1.61$. The cross enclosed 
in a square marks the position of CXOCY J033716.7-050153 after correcting for optical
obscuration, while the arrow indicates the displacement caused by this correction. The uncertainty of the X-ray flux measurement
of CXOCY J033716.7-050153 is $\sim 30\%$ and is not included for clarity.}
\end{figure}

\clearpage

\begin{deluxetable}{cc}
\tablecaption{Properties of CXOCY J033716.7-050153\label{tab1}}
\tablecolumns{2}
\tablewidth{0pt}
\startdata
\tableline\tableline
Parameter & Value \\
\tableline
RA (J2000) (X-ray) & $03^h37^m16.6s$ \\
Dec (J2000)(X-ray) & $-05^\circ 01'54.3''$ \\ 
RA (J2000) (optical)& $03^h37^m16.6^s$ \\
Dec (J2000)(optical)& $-05^\circ 01'53.7''$ \\
Galactic N$_H$\tablenotemark{a} & $4.8\times 10^{20}$ cm$^{-2}$\\ 
$z$ & $4.61\pm 0.01$\\
$\alpha_{OX}$ & $-1.16\pm 0.16$\\ 
B mag\tablenotemark{b}& $>26.5$\\
V mag\tablenotemark{b} & 25.76$\pm 0.39$\\
R mag\tablenotemark{b} & 24.01$\pm 0.08$\\
I mag (CTIO)\tablenotemark{b} & 23.95$\pm 0.23$\\
I mag (VLT)\tablenotemark{b} & 23.66$\pm 0.07$\\
$M_I$\tablenotemark{c} & -22.14$\pm 0.15$\\
$AB_{1450(1+z)}$ & 23.82\\
$M_{1450(1+z)}$ & -22.45\\
$f_X$ (0.5-8 keV) & $3.5 \times 10^{-15}$ ergs s$^{-1}$ cm$^{-2}$\\
$f_X$ (0.5-2 keV) & $1.8 \times 10^{-15}$ ergs s$^{-1}$ cm$^{-2}$\\
$L_X$ (0.5-2 keV) & $4.5 \times 10^{44}$ ergs s$^{-1}$\\
\enddata
\tablenotetext{a}{calculated using HEASARC tool nh}
\tablenotetext{b}{AB magnitudes in a $1.4\times$ FWHM aperture}
\tablenotetext{c}{Rest Frame $I$ band AB magnitude, assuming the SDSS average quasar spectrum} 
\end{deluxetable}

\end{document}